Kamil Koniuch[a]*, Sabina Baraković[b] , Jasmina Baraković Husić[c] Katrien De Moor[d] , Lucjan Janowski[a] , Michał Wierzchoń[e,f,g]

[a] *Institute of Communication Technologies, AGH University of Science and Technology, Cracow, Poland;* [b] *Faculty of Traffic and Communications, University of Sarajevo, Sarajevo, Bosnia and Herzegovina Zmaja od Bosne bb, 71000 Sarajevo, Bosnia and Herzegovina;* [c] *Faculty of Electrical Engineering, University of Sarajevo, Sarajevo, Bosnia and Herzegovina Zmaja od Bosne bb, 71000 Sarajevo, Bosnia and Herzegovina;* [d] *Department of Information Security and Communication Technology, Norwegian University of Science and Technology, Trondheim, Norway;* [e] *Consciousness Lab, Institute of Psychology, Jagiellonian University, Kraków, Poland;* [f] *Centre for Brain Research, Jagiellonian University, Kraków, Poland;* [g] *Jagiellonian Human-Centered Artificial Intelligence Laboratory, Jagiellonian University, Kraków, Poland*

*corresponding author koniuch@agh.edu.pl,*

Kamil Koniuch is a PhD student at the Institute of Communication Technologies, AGH University of Science and Technology since 2020. He is a member of AGH Video Quality of Experience (QoE) team. His research focuses on the ecological validity of QoE measurements. He received his master's degree in psychology in 2019. He is a graduate of the Institute of Psychology, Jagiellonian University.

ORCID: 000-0002-8243-7155

Sabina Baraković, is an Associate Professor at the University of Sarajevo and Expert Adviser at the Ministry of Security of Bosnia and Herzegovina. She is a Co-founder of a research laboratory Little Mama Labs. She is/was principal investigator on many national and international projects (Horizon2020, NATO, ITU, Erasmus+, CoE, OSCE, etc.) and has co-operators throughout the world (e.g., EU member countries, SEE region countries, Qatar, etc.). Her research interests include multidimensional QoE modelling of multimedia, mobile Web-based applications, and unified communications, cybersecurity in smart environments and IoT/WoT, QoS/QoE/QoL management, etc.

ORCID: 0000-0001-8121-9765

Jasmina Baraković Husić, is a Full Professor at the University of Sarajevo. She is employed by BH Telecom, Joint Stock Company, Sarajevo as a specialist in the Department for Planning and Development of Technologies and Services. She is co-founder of a research laboratory Little Mama Labs. She is/was principal investigator on many national and international projects (NATO, Erasmus+, Ministry of Education, RAI, etc.) and has co-operators throughout the world (e.g., EU member countries, SEE region countries, etc.). Her research interests include variety of topics in networked multimedia services, QoS/QoE/QoL management, 5G-IoT communications.

ORCID: *0000-0001-6119-6447*

Katrien De Moor is associate professor at the department of Information Security and Communication Technology at NTNU since 2019. After obtaining her PhD degree in Social Sciences from Ghent University (2012), she did a Marie Curie postdoctoral fellowship at NTNU. She is co-Editor-in-Chief of the multidisciplinary journal "Quality and User Experience" (Springer) since 2018 and one of the 20 founding members of the Young Academy of Norway. Her main research interests and activities are linked to human-technology experiences (e.g., video-based and immersive experiences...) and human-technology behavior, related methodological challenges (inclusion and user diversity, ecological validity, ...), as well ethical implications (e.g., agency, power dynamics in design processes, ecological footprint of ICT).

ORCID: 0000-0002-8752-3351

Lucjan Janowski received the PhD degree in telecommunications from the AGH University of Science and Technology, Krakow, Poland, in 2006. In 2007, he was a Postdoctoral Researcher with the Laboratory for Analysis and Architecture of Systems, Centre National de la Recherche Scientifique, Paris, France. From 2010 to 2011, he was a Postdoctoral Researcher with the University of Geneva, Geneva, Switzerland. From 2014 to 2015, he was a Postdoctoral Researcher with the Telecommunications Research Center Vienna, Vienna, Austria. He is currently an Assistant Professor with the Institute of Communication Technologies, AGH University of Science and Technology. His research interests include statistics and probabilistic modeling of subjects and subjective rates used in QoE evaluation.

ORCID: 0000-0002-3151-2944

Michał Wierzchoń is a Professor of Social Sciences. He currently serves as the director of the Centre for Brain Research and the Vice-Dean for the Scientific Affairs of the Faculty of Philosophy. He is also the Founder and the Head of the Consciousness Lab (C-Lab). He has been recently nominated as the President-Elect of the European Society for Cognitive Psychology (2023-2024). His research focuses on the neuronal and cognitive mechanisms of consciousness, including conscious perception, memory and self-awareness. He uses various research methods, such as behavioural, neuroimaging, qualitative studies, and computational modelling. In the context of applied studies, his work focuses on the development of sensory substitution systems for blinds and the evaluation of media quality.

ORCID: 0000-0002-7347-2696

# Top-down and bottom-up approaches to video Quality of Experience studies; overview and proposal of a new model

Modern video streaming services require quality assurance of the presented audiovisual material. Quality assurance mechanisms allow streaming platforms to provide quality levels that are considered sufficient to yield user satisfaction, with the least possible amount of data transferred. A variety of measures and approaches have been developed to control video quality, e.g., by adapting it to network conditions. These include objective matrices of the quality and thresholds identified by means of subjective perceptual judgments. The former group of matrices has recently gained the attention of (multi)media researchers. They call this area of study "Quality of Experience" (QoE). In this paper, we present a review of QoE's theoretical models together with a discussion of their properties and implications for the field. We argue that most of them represent the bottom-up approach to modeling. Such models focus on describing as many variables as possible, but with a limited ability to investigate the causal relationship between them; therefore, the applicability of the findings in practice is limited. To advance the field, we therefore propose a structural, top-down model of video QoE that describes causal relationships among variables. We hope that our framework will facilitate designing comparable experiments in the domain.

Subject classification codes: include these here if the journal requires them

## Introduction

The quality of multimedia is a subject of interest for many practitioners and researchers in various domains, including telecommunications engineering, speech, audio and video processing, psychophysics, human-computer interaction, psychology, ergonomics, human factors research, and innovation and economics (Raake & Egger, 2014). Quality of Experience (QoE) as a part of the broader field of multimedia research focuses on the subjective quality perception of a wide range of multimedia services. More concretely, QoE research aims to identify factors and features that are key to enabling or inhibiting good user experiences and to optimize them so that the produced quality levels are perceived as high, able to satisfy users' expectations and contribute to users' willingness to reuse the service in the future (Möller & Raake, 2014).

For many years, the literature was fragmented with respect to conceptual interpretations and definitions of QoE (see, e.g., (De Moor, 2012) for an overview). However, a more recent community-level effort as part of the Qualinet project resulted in a new, holistic definition (Brunnström, et al., 2013). According to this definition, QoE is "*the degree of delight or annoyance of the user of an application or service. It results from the fulfillment of his or her expectations with respect to the utility and/or enjoyment of the application or service in the light of the user's personality and current state*". This definition is very broad and covers multimedia in general. In practice, experts in different types of multimedia (e.g., web browsing (Baraković & Skorin-Kapov, 2015; Baraković & Skorin-Kapov, 2017; Baraković & Skorin-Kapov, 2017), unified communications (Baraković Husić, et al., 2020; Barakovic Husic, Barakovic, Krejcar, & Maresova, 2021), sound (Ragano, Benetos, & Hines, 2019), and gaming (Laghari, et al., 2019)) study specific multimedia services separately. This is due to the fact that different use cases and types of services require an understanding of the key factors that play a role and that can be taken into account to improve QoE. These key factors tend to be application and use case-specific, and hence lead to a diverse set of variables to be operationalized, captured, and understood.

For this reason, describing models specific to one multimedia type might be beneficial, as they can provide actionable insights for a certain type of service. In this paper, motivated by both the relevance of insights to industry players in the video streaming ecosystem and the ever-increasing share of Internet traffic that is due to video-based services, we will focus only on QoE in the context of video streaming. Our aim is, however, to refine and adjust the model presented in this paper so that it also becomes applicable to other types of multimedia services in the future.

In the context of video quality, modern services use insights from research for the improvement and optimization of user experience. For example, user assessments helped develop video multimethod assessment fusion (VMAF), an objective full-reference metric that predicts subjective ratings automatically (Li, Aaron, Katsavounidis, Moorthy, & Manohara, 2016). These types of metrics are used in the development of new video compression algorithms that aim for the reduction of energy consumption. As a result, quality-aware optimization may help to increase the sustainability of video-on-demand services (Lee, Lee, & Song, 2019). Thus, QoE studies can be part of the growing research on the influence of video services on the natural environment (Ejembi & Bhatti, 2015; Batmunkh, 2022; Chandaria, Hunter, & Williams, 2011).

In this context, estimating aspects relating to the sustainability of services (e.g., energy efficiency) is possible because the QoE concept not only describes the user interaction with the service but also provides a framework for optimization. Specifically, the latter is possible due to standardized measurement and test protocols, which allow researchers to measure subjective judgments of quality. In traditional standardized experiments on video QoE, short, soundless video clips are used as stimuli (ITU-R, 2020). After the presentation of each stimulus, a single subjective quality judgment is collected. This procedure is repeated for dozens of videos, and the same content is presented multiple times. A key criticism of this approach is that this research paradigm is simplistic and limits the scope of the investigated variables. In effect, it offers reliable results but provides little insight into how people perceive video quality in their natural environment, where users are influenced by many factors besides the quality of video compression. Thus, metrics stemming from data sets gathered with this paradigm might misjudge everyday user experience, including associated cognitive processing and implications for user behavior.

To meet this challenge, researchers have investigated influential factors (IFs) of QoE. IFs are defined as "*any characteristic of a user, system, service, application, or context whose actual state or setting may have an influence on the Quality of Experience for the user*" (Brunnström, et al., 2013). In the context of video services, there is a plethora of variables that could potentially influence the user. Furthermore, researchers have investigated perceptual dimensions (PDs) or "perceptual features" of QoE (Baraković Husić & Baraković, 2022). These variables are defined as "*perceivable, recognizable, and nameable characteristics of the individual's experience of a service which contributes to its quality*" (Jekosch, 2005). The concept of PDs highlights the differences between objective changes in stimuli or context and the subjective perception of those changes.

Consequently, QoE can be considered a multidimensional concept with many variables that need to be considered when designing an experiment. As human perception is a key component of QoE, many of those variables are latent. A strong theoretical background is therefore necessary to provide comparable operationalization and experimental procedures. Accordingly, a series of extensive theoretical models have been proposed to describe the complex character of QoE (Brunnström, et al., 2013; Raake & Egger, 2014; Möller, Wältermann, & Garcia, 2014; Reichl, et al., 2015; Robitza, Schönfellner, & Raake, 2016; Schmitt, Bulterman, & Cesar, 2018; Reiter, et al., 2014); (Baraković Husić & Baraković, 2022; Geerts, et al., 2010; Egger, Reichl, & Schoenenberg, 2014). They comprise many variables and provide a taxonomy of IFs and PDs.

To investigate which of these IFs are most important for QoE, researchers use exploratory study designs that are often based on self-report approaches, such as questionnaires (Baraković Husić & Baraković, 2022; Zhu, Heynderickx, & Redi, 2015). With this method, researchers are able to measure a vast number of variables in a natural context (Staelens, et al., 2010). This approach to the study of IFs and PDs can be described as bottom-up (see Figure 1). In this method, the goal is to investigate as many variables as possible.

Nevertheless, research designed with the bottom-up approach often does not allow for the investigation of the causal relationship between variables in the model. Without this information, models can miscalculate the importance of certain IFs or PDs due to the misinterpretation of confounding variables (McElreath, 2020). A top-down approach is more suitable to investigate how information about causal relationships between variables can change the estimation of their influence (Figure 1). This method allows for the investigation of not only the influence of the predictors but also their interrelations in the form of a graph. Moreover, the use of diagrams describing the relationships between variables clarifies the experimental design process. With a visual representation of assumptions and variables, it is easier to find adequate statistical methods and to communicate the conclusions with other researchers. In effect, it is easier to design experiments that are comparable with previous studies. Although they thus have clear advantages, to the best of our knowledge, there is a limited number of QoE studies that use top-down approaches, i.e., models that start with a predefined list of variables and test their influence.

Given the above, this paper aims to present a video QoE model that can be verified with experimental data. Moreover, our model provides a structure for the investigation of causal relations between IFs and PDs. Finally, we hope that our model will facilitate the discussion by providing a clear taxonomy in QoE studies. In this paper, we will focus on the video QoE, but our model can be adjusted to other use cases. The rest of this paper is organized as follows: In the next section, these two modeling approaches and their advantages and disadvantages are compared and discussed in detail. Furthermore, we present the components of our model with their operationalization. Finally, we present a generalized video QoE model describing the causal relation between variables and discuss the implications for video QoE research.

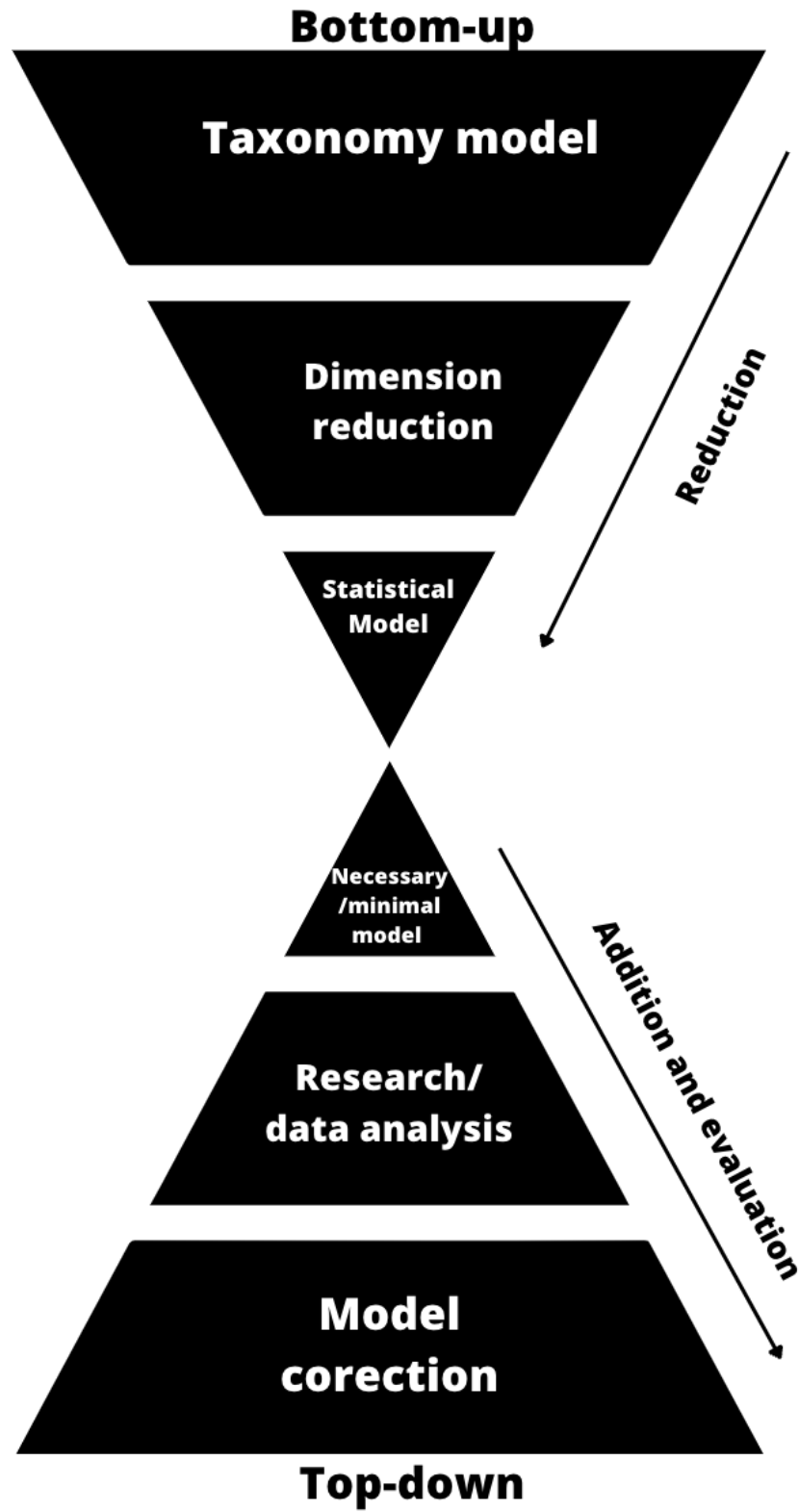

*Figure 1. Modeling approaches. Figure sizes represent the number of variables in the model and arrows*

**Top-down and bottom-up approaches**

To visualize the differences between previously published QoE models and our model,

we describe them in terms of *bottom-up* and *top-down* approaches (Figure 1). These terms are used to characterize models in various domains, from cognitive processes (e.g., attention; see (Connor, Egeth, & Yantis, 2004)) to the causal direction of subjective wellbeing (Headey, Veenhoven, & Weari, 2005). In this paper, we use these terms to underscore differences between approaches in the number of variables taken into account and the type of their analysis. The trade-off between the complexity of models and the amount of information they provide is one of the key problems researchers must address. A good model predicts as much as possible with as few assumptions and variables as possible (Simon, 2001; Gabaix & Laibson, 2008). This property of the model is called parsimony. Only the most important variables must be determined to achieve parsimony. To identify the key components of the model, one can start with as many variables as possible and search for the most important ones. We call this a bottom-up approach (see Figure 1). On the other hand, one can start building the model by defining the minimum number of variables that are necessary and sufficient to describe the phenomenon. We call this a top-down approach. We discuss these approaches below in greater detail.

### *Bottom-up approach*

The bottom-up approach is typically used for the *exploration* of data sets. The main goal is to find the most important variables or patterns and, if possible, to identify the latent structure of the phenomenon. Models comprising the most important variables are outcomes of this method. This type of research is often used as a source of hypotheses for following experimental designs. Below, we present a short overview of the bottom-up methods used in QoE research.

Over the years, a vast number of statistical methods have been developed to provide insight into data sets. In the context of QoE research, the statistical analysis method group works on the implementation of new methods in the field (VQEG, 2022).

Methods such as principal component analysis (PCA) (Ketykó, De Moor, Joseph, Martens, & De Marez, 2010; Msakni & Youssef, 2016; Baraković Husić & Baraković, 2022) and factor analysis (FA)(e.g. (Özer, Argan, & Argan, 2013; Ende, 2015)) are typically used to investigate latent structures in data. Analysis of variance, covariance, and correlations can also be used to investigate data patterns. Recently, machine learning and network analysis approaches have become popular solutions for analyzing complex, nonlinear relations (Youssef, Mellouk, Afif, & Tabbane, 2016). All the abovementioned methods can be used on data sets comprising many variables and factors. Nevertheless, it is impossible to include all the potential IFs and PDs in one experimental setup. Thus, to provide that variety, questionnaires are often used (Baraković Husić & Baraković, 2022; Ketykó, De Moor, Joseph, Martens, & De Marez, 2010; Msakni & Youssef, 2016; De Pessemier, Martens, & Joseph, 2013).

| Study | Analyse method | Data type | Sample |
|---|---|---|---|
| (Ketykó, De Moor, Joseph, Martens, & De Marez, 2010) | PCA and correlation analysis | Questionnaire, Experience Sampling Method | 19 participants |
| (Msakni & Youssef, 2016) | PCA and correlation analysis | Emotional Scale, ACR scale | 72 participants |
| (Youssef, Mellouk, Afif, & Tabbane, 2016) | PCA Machine learning | ACR scale | 45 participants |
| (Baraković Husić & Baraković, 2022) | PCA Multiple Linear | Questionnaire, ACR scale | 233 participants |

| | Regression (MLR) | | |
|---|---|---|---|
| (Özer, Argan, & Argan, 2013) | PCA Confirmatory Factor Analysis (CFA) | Questionnaire | 1000 participants |
| (Ende, 2015) | Exploratory Factor Analysis CFA T-test Correlation analysis | Questionnaire, ACR scale | Exploration: 107 participants Confirmation: 100 participants |
| (De Pessemier, Martens, & Joseph, 2013) | MLR | Experience Sampling Diary Questionnaire | 29 participants |

Besides statistical solutions, typical for an exploratory study, this approach can also be used with qualitative research. Analysis of interviews and open questions in questionnaires are a rich source of information that must be condensed and structured by researchers (Ickin, et al., 2012; Staelens, et al., 2010). Similarly, bibliometric studies provide valuable insights into a vast amount of previous results and conclusions (Zhao, Liu, & Chen, 2016; Vega, Perra, De Turck, & Liotta, 2018; Maia, Yehia, & de Errico, 2015; Skorin-Kapov, Varela, Hoßfeld, & Chen, 2018; Alreshoodi & Woods, 2013; Alreshoodi & Woods, 2013; Juluri, Tamarapalli, & Medhi, 2016). Moreover, mixed methods, such as open profiling of quality (OPQ) (Strohmeier, Jumisko-Pyykkö, Kunze, & Bici, 2011), can provide quantitative results based on qualitative data. In other cases, mixed-method approaches have been used to deepen the understanding of quantitative data by means of a qualitative follow-up (see, e.g., (De Moor, Borge, & Heegaard, 2019; Øie, Koniuch, Cieplińska, & De Moor, 2021)). All the abovementioned methods share a similar goal. Their aim is to better understand the collected data. This type of insight is especially useful in the study of complex phenomena such as QoE.

Nevertheless, this approach has some limitations. Bottom-up methods rarely provide information about the causal relationship between variables such as IFs or PDs of QoE. In effect, it is easy to misjudge the importance of some variables. Moreover, it is possible that there are confounding variables that can influence both the predictor and the predicted value. Another source of interpretation error could be the character of the analyzed data. All measurements are contaminated with some unknown measure error. In the case of self-descriptive data, scientists need to be especially cautious. There are numerous effects that can influence participants' responses. Alternatively, the reduction of variables in the model can be achieved with a simplified experimental design. Controlled laboratory experiments aim for this type of reduction (ITU-R, 2020). In this approach, researchers focus mostly on the manipulation of technical factors to better predict their influence on perceived video quality. This methodology was proven to be useful for building quality metrics such as VMAF (Li, Aaron, Katsavounidis, Moorthy, & Manohara, 2016) and for the assessment of new compression algorithms. Nevertheless, the ecological validity of the prediction of VMAF remains undetermined. The discussed experiments are far from the everyday experience of users, and metrics built with this approach can miscalculate user satisfaction.

To sum up, bottom-up modeling is dominant in QoE research. There are numerous theoretical models that try to gather all potential influencing factors and perceptual dimensions of (Brunnström, et al., 2013; Raake & Egger, 2014; Möller, Wältermann, & Garcia, 2014; Reichl, et al., 2015; Robitza, Schönfellner, & Raake, 2016; Schmitt, Bulterman, & Cesar, 2018; Reiter, et al., 2014); (Baraković Husić & Baraković, 2022; Geerts, et al., 2010; Egger, Reichl, & Schoenenberg, 2014). They highlight the complexity of human perception and provide a taxonomy for IFs and PDs. These models have inspired a number of studies investigating the most important predictors of QoE; in particular, system level and technical IFs have received a lot of attention. Nevertheless, causal relations between IFs, PDs, and QoE are still not well investigated, despite pleas in the literature to address them (see, e.g., (Reiter, et al., 2014)). For that purpose, top-down approaches can be more useful.

### *Top-down approach*

A top-down approach is a theory-driven approach that is used for model comparison. Thanks to this approach, it is possible to decide between alternative theoretical and statistical models. With this line of action, abstract concepts such as theories can be verified with observational or experimental data. In effect, models can be developed by adding new insights from experiments. Most importantly, this approach allows for the verification of conclusions provided by data-driven models. Additionally, with this approach, it is possible to investigate causal relationships between variables and determine confounding variables.

Currently, statistical methods that can be used for the verification of theoretical models are being intensively developed. The approaches differ in technical details but share similar assumptions. All of the methods are based on graphs that represent the relationship between investigated variables, so-called path analysis. Consequently, with path analysis, it is not only possible to add new predictors but also to determine their causal relationships. They can verify models by measuring how well they fit the data gathered in the experiments. The most popular path analysis model is confirmatory factor analysis (CFA), which is a part of structural equation modeling and causal structural models.

The main shortcoming of this approach is the limited number of variables that can be investigated. The more units a graph includes, the more potential connections between them must be represented and thus analyzed. Moreover, it is harder to draw conclusions from statistics provided by model comparison. For example, a more *informative* model can have a worse statistical fit. Lastly, experiments that can provide adequate data are complex and must be run on large samples.

Nevertheless, to answer modern QoE problems, top-down modeling is necessary. The lack of ecological validity of standardized experiments requires the measurement of a large number of factors and more complicated manipulations. Understanding not only testers' ratings but consumers' behavior in general requires a better understanding of the relationship between variables in data sets. Moreover, without investigating causal relations between variables influencing consumer behavior, researchers are vulnerable to confounding factors.

However, to use the theory-driven approach in data modeling, a simplification of current QoE models is required. We therefore propose an additive approach to QoE modeling. In the next section, we present a graph model of video QoE comprising a minimal number of the necessary, most important variables. This model can be developed with further IFs and PDs, maintaining the core of the model intact. In other words, we present a general model that can be adjusted by researchers to investigate specific contexts. While we propose the model for video QoE, it is possible to adjust it for other use cases.

**Video QoE model based on the path diagram**

We based our model on the interpretation of the first part of the general definition of QoE: "*Quality of Experience is the degree of delight or annoyance of the user of an application or service.*" For this reason, we did not represent QoE as a separate unit in the Figure 2. Instead, we included a "delight or annoyance" unit as an outcome of the interaction of variables typical for video service experience. In real life, it is hard to imagine a scenario in which the delight or annoyance of the user is not generated by the video content. Depending on internet efficacy, this content-dependent experience *might be* moderated by drops in video quality. With this reasoning, we built a path model of QoE.

Following previous QoE models, we distinguished IFs and PDs as predictors of general user satisfaction. IFs are defined as "*any characteristic of a user, system, service, application, or context whose actual state or setting may have an influence on the Quality of Experience for the user*" (Reiter, et al., 2014). PDs refer to "*perceivable, recognizable, and nameable characteristics of the individual's experience of a service which contributes to its quality.*" Moreover, we described these variables as latent variables and provided examples of their measurements. Below, we briefly present the operationalization of each variable in our model.

*Component operationalization*

*QoS*

Quality of Service (QoS) is a measure of the overall performance of a network. It is often used to describe the ability of a network to provide a consistent level of service to its users. ITU-T defines QoS as "*The totality of characteristics of a telecommunications service that bear on its ability to satisfy stated and implied needs of the user of the service*" (Recommendation, 2008). QoS can be expressed in terms of performance indicators such as throughput, latency, jitter, and packet loss. Special metrics for QoS assessment are constantly being developed. There are also models describing the influence of QoS on QoE (Fiedler, Hossfeld, & Tran-Gia, 2010; Chen, Wu, & Zhang, 2014; Kim & Choi, 2014). QoS is a well-known QoE IF.

*Content*

Content characteristics are multidimensional. Depending on the research question, different variables might be taken into account to analyze its influence. Due to its complex character, content can be identified as a system or a context IF. For example, characteristics such as motion, number of details, brightness, and computation complexity are well recognized in QoE studies (Song & Tjondronegoro, 2014; Duanmu, Ma, & Wang, 2018; Khan, Sun, & Ifeachor, 2011). They are categorized as system factors influencing QoE. On the other hand, the content type chosen by the user can be classified as a context IF and can be used for predicting user satisfaction (Agboma & Liotta, 2012). Moreover, influential service providers such as YouTube have their own metrics for classifying and quantifying content. One of the crucial statistics for distinguishing videos in the service is "engagement", operationalized as the mean percentage of watching time. It can be used for the operationalization of PDs of the content such as perceived engagement or interest in the video. Furthermore, self-description methods can be used for assessment of participant level of interest (Song, Yang, Zhou, Wan, & Wu, 2016) or motivation (Kobayashi, Masuda, & Hayashi, 2012). Moreover, there are standardized data sets of visual stimuli that can be used to evaluate the influence of emotions evoked by content (Marchewka, Żurawski, Jednoróg, &

Grabowska, 2014; Di Crosta, et al., 2020; Baveye, Dellandrea, Chamaret, & Chen, 2015; Baveye, Dellandrea, Chamaret, & Chen, 2015).

*Quality of multimedia signal*

In our model, quality of multimedia signal represents the objective properties of visual stimuli. In the context of video streaming, it is video displayed on the user's device. There are many methods for assessing the quality of video, some of which are objective and some subjective. Thus, we propose a division into the objective quality of media signal (QoMS) and its PD – the perceived QoMS described below. In this approach, QoMS is the IF that represents the quality of reproduction of the source signal (content). It has properties of the source content that are moderated by the efficacy of the network and user hardware. It can be assessed with objective metrics such as signal-to-noise ratio or VMAF (García, Gortázar, Gallego, & Hines, 2020; García, López-Fernández, Gortázar, & Gallego, 2019; Li, Aaron, Katsavounidis, Moorthy, & Manohara, 2016). In current models, (Brunnström, et al., 2013; Raake & Egger, 2014; Robitza, Schönfellner, & Raake, 2016), QoMS is described as the physical representation of the signal.

*Perceived Quality of multimedia signal*

We use the term "perceived QoMS" (PQoMS) to emphasize the role of perception in video QoE studies where subjective assessments of quality made by users are in the spotlight. Typically, researchers estimate QoMS with a 5-point Absolute Category Rating (ACR) scale (ITUT, 2016). Additionally, there is a new effort to build matrices that take into account both QoMS and PQoMS to predict user satisfaction (Li, Aaron, Katsavounidis, Moorthy, & Manohara, 2016). In proposed models (Raake & Egger, 2014; Brunnström, et al., 2013), these descriptions are the outcome of the quality formation process.

*Degree of delight or annoyance*

We assume that the user's state of delight or annoyance (DoA) is the outcome of both quality and content properties. According to the general definition, this is, in fact, the measure of QoE. It can be assessed, for example, with an adapted Differential Emotions Scale (De Moor, Quintero, Strohmeier, & Raake, 2013). In the natural context, both technical (Nam, Kim, & Schulzrinne, 2016) and content (Laiche, Ben Letaifa, Elloumi i Aguili, 2021) related factors may lead to a change in the QoE. This might cause a shift in user behavior (Robitza, Schönfellner, & Raake, 2016).

*Behavior*

Depending on the scope of the study, behavior might be a short-term reaction to quality-related events (Robitza & Raake, 2016; Fogelberg, 2020; Laiche, Ben Letaifa i Aguili, 2021), habit evaluation (Robitza, et al., 2020), or even consumer attitude (Reichl, et al., 2015; Perkis, Reichl, & Beker, 2014) predictors. Generally in the retail context, pleasure and arousal are good predictors of approach–avoidance behavior (Kenhove & Desrumaux, 1997). As long-term behavior toward network providers might be influenced by a set of additional important variables (e.g., pricing), we focus on short-term behavior. In our model, behavior is an outcome of DoA and can be observed as a change in interaction with service (e.g., change of the video).

**Relationships between variables and model assumptions**

We described our generalized QoE model in the form of a diagram (see Figure 2) with causal paths between variables. Following the causal path analysis approach, arrows represent the assumption that variable A may be a direct cause of B. For example, in Figure 2, DoA *might be* caused by perceived QoMS and content. Additionally, arrow representation does not imply that the process is simple. Relationships between A and B can be complex, multi-staged, and nonlinear, what we assume is the direction of the relationship. Moreover, by drawing an arrow from A to B, we do not assume that A is the only thing that causes B. In fact, this approach assumes that every variable can be influenced by unspecified factors not represented in a graph. This enables including measurement error and the influence of unknown factors in the model, as well as adding new variables in future updates of the model. However, if we know that units in the model have a common cause, this must be expressed in the graph.

On the other hand, the absence of an arrow represents a stronger assumption, namely, the lack of relationship between variables represented by nodes (Paulewicz, Siedlecka, & Koculak, 2020). For example in Figure 2, the amount of delight or irritation does not change the QoMS; as such, the direction of influence seems impossible.

In our model, we described the physical representation of the video signal (QoMS) as an outcome of the interaction between network efficacy (QoS) and Content. We assumed that from the user's perspective, there is no relationship between QoS and Content other than QoMS. In other words, if network efficacy has some influence on video content, it can only be observed via the QoMS. In consequence, QoS cannot directly influence DoA or behavior. Users must see the change in QoMS to conclude there are network efficiency problems. Furthermore, the arrow from QoMS to Perceived QoMS represents human perception. In fact, this process could be described using quality formation models such as (Raake & Egger, 2014; Brunnström, et al., 2013). Moreover, it could be influenced by cognitive factors such as visual sensitivity (Banitalebi-Dehkordi, Ebrahimi-Moghadam, Khademi, & Hadizadeh, 2019). In addition, we cannot assume that the perception of quality is not moderated by DoA. That is why the arrow between PQoMS and DoA is bidirectional.

Most importantly, we assume that in real-life scenarios, DoA is a function of both content and PQoMS. Users might react differently to the same QoMS depending on the type of content.

Finally, users' behavior is the consequence of that DoA (Seufert, Wassermann, & Casas, 2019; Hu, et al., 2020). One can be dissatisfied both due to the content and the PQoMS. This can result in behavior directed to enhance network efficacy, change of content, compensation, or abandonment of activity.

The model presented in this paper is general. This means that we treat it as a framework that can be specified for particular use cases and experimental setups. As already mentioned, in such cases, the model can be extended with additional variables. This procedure requires the operationalization and inclusion of new units in the graph. For example, if we want to add participant interest, we can place it on the arrow from content to DoA. In that case, we assume that interest is caused by content but not by other variables. Moreover, content can only influence general satisfaction by the level of interest. If one's hypothesis is that interest does not always influence the causal path between content and DoA, interest should be included with an additional path. Another hypothesis might be that the perceived QoMS is influenced by visual sensitivity. In such a scenario, we must add visual sensitivity as a new unit outside the graph and draw the line to perceived QoP. Other influential factors from the model (Reiter, et al., 2014) could be added in a similar manner.

Based on this causal structure, we can propose an operational definition of QoE limited to video services. QoE is the amount of behaviorally relevant user DoA toward a video service evoked by content and moderated by the perceived quality of the video.

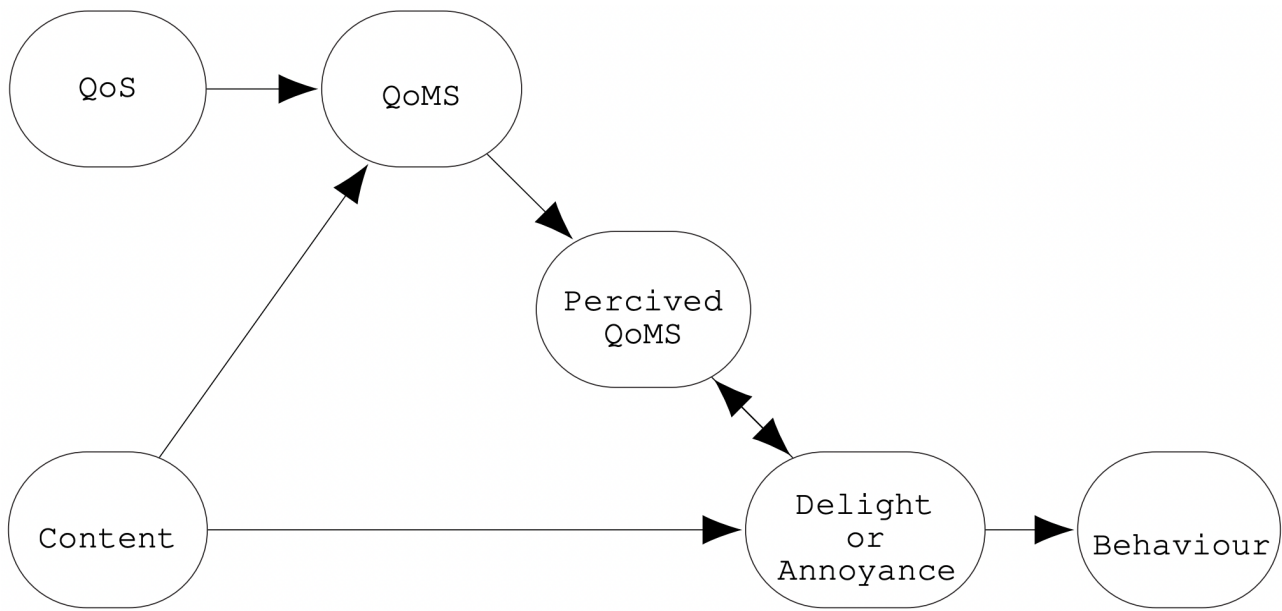

*Figure 2 path model of video QoE*

## Implications and future directions

As presented in the section above, our model can be used to operationalize new variables, build hypotheses, and investigate assumptions of experimental setups. Due to its causal structure and clear operationalization of variables, our model can be verified with experimental data. Moreover, our assumptions, represented by arrows, can be verified. Path analysis such as structural equation modeling or causal structural models can be used for validation of our assumptions. However, the number of units included in the model is limited. Future development of the model should comprise validation and inclusion of other factors known from bottom-up QoE research. In this way, more complex models could be built in an additive manner. A comparison of those models will require a new experimental design that will include a broader spectrum of variables. In effect, we hope that our model will help not only to distinguish which compression method is better but also provide answers to more complex questions such as, "How important is quality?" and "Why do people stop watching videos?" Is it, e.g., due to the lack of content attractiveness or because the video quality is insufficient? We are convinced that answers to those questions will help to provide more sustainable solutions for video streaming in the future. Although our model describes video QoE, it can also be adjusted to different multimedia in the future. Finally, we hope that our model will be a source of inspiration for new hypotheses and experimental studies in the QoE domain. Application of our model may provide structure for the discussion of QoE and lead to the creation of comparable, replicable paradigms. This new structure could possibly facilitate dialogue among specialists from different domains relative to QoE.

Acknowledgements, avoiding identifying any of the authors prior to peer review

Disclosure Statement

No potential conflict of interest was reported by the author(s).

Funding

The research leading to these results has received funding from the Norwegian

Financial Mechanism 2014-2021 under project 2019/34/H/ST6/00599.